\documentclass[nofootinbib,aps,superscriptaddress,preprint,floatfix]{revtex4}

\usepackage{graphicx,epsf}

\usepackage{bm}

\def\bs{$/\!\!/$}
\def\to{\rightarrow}
\newcommand{\nc}{\newcommand}
\nc{\beq}{\begin{equation}}
\nc{\eeq}{\end{equation}}
\nc{\barray}{\begin{eqnarray}}
\nc{\earray}{\end{eqnarray}}
\nc{\barrayn}{\begin{eqnarray*}}
\nc{\earrayn}{\end{eqnarray*}}
\nc{\bcenter}{\begin{center}}
\nc{\ecenter}{\end{center}}
\nc{\ket}[1]{| #1 \rangle}
\nc{\bra}[1]{\langle #1 |}
\nc{\mc}{\mathcal}
\nc{\er}[1]{(\ref{eq:#1})}
\nc{\onehalf}{\frac{1}{2}}
\nc{\partialbar}{\bar{\partial}}
\nc{\psit}{\widetilde{\psi}}
\nc{\Tr}{\mbox{Tr}}
\nc{\hc}{\mbox{H.c.}}
\nc{\ev}{\;\mathrm{eV}}
\nc{\mev}{\;\mathrm{MeV}}
\nc{\gev}{\;\mathrm{GeV}}
\nc{\tev}{\;\mathrm{TeV}}

\def\ttbar{$t \bar t$}
\newcommand{\gsim}{\lower.7ex\hbox{$\;\stackrel{\textstyle>}{\sim}\;$}}
\newcommand{\lsim}{\lower.7ex\hbox{$\;\stackrel{\textstyle<}{\sim}\;$}}

\newcommand{\afbtt}{A_{\rm FB}^{t\bar t}}
\newcommand{\afbl}{A_{\rm FB}^{\ell}}
\newcommand{\afbll}{A_{\rm FB}^{\ell \ell}}
\newcommand{\dzero}{$D\O$}

\begin{document}
%\setlength{\topmargin}{+0.3in}
%\voffset 30pt
%\vspace{3.0cm}
%\preprint{\vbox

%\vspace*{2cm}

\preprint{CERN-PH-TH/2011-248, LPT-ORSAY 11-666}

\title{\bf \Large Spinning the Top}

\author{\large   Adam Falkowski}
\affiliation{Laboratoire de Physique Th\'eorique d'Orsay, UMR8627--CNRS,\\ Universit\'e Paris--Sud, 91405 Orsay, France}

\author{\large Gilad Perez}
\affiliation{CERN, Theory Division, CHÐ1211 Geneva 23, Switzerland}
\affiliation{
Department of Particle Physics and Astrophysics, Weizmann Institute of Science, Rehovot 76100, Israel
}

\author{\large  Martin Schmaltz}
\affiliation{Physics Department, Boston University, Boston, MA 02215
                            }

\vspace{1cm}

\begin{abstract} 
We propose to measure the {\em threshold lepton asymmetry}, that is the forward-backward asymmetry of the charged lepton in \ttbar\ events near the production threshold.   
At threshold top quark pairs are produced in an s-wave. Angular momentum conservation then implies that the top spins equal the spin of the initial state which -- in the case of quarks -- is uniquely fixed by the chirality of the initial quarks. Thus measuring final state top spins determines the chirality of the quarks which produced them. Information about the top spins can be extracted by measuring the angular distribution of the charged lepton in semileptonic or dileptonic decays of the top pair. One such distribution, the threshold lepton asymmetry, vanishes in tree-level QCD but is non-zero if new physics modifies the relative contribution of right-handed and left-handed quarks to top pair production. This is interesting because realistic models addressing the anomalous  \ttbar\ asymmetry have chiral couplings to light quarks. Models with identical \ttbar\ asymmetries at the Tevatron can be distinguished by their threshold lepton asymmetries which range between plus and minus 25\% in realistic models.   
\end{abstract}

\def\thepage{{}}
\maketitle
\def\thepage{\arabic{page}}

%%%%%%%%%%%%%%%%%%%%%%%%%%%%%%%%%%%%%%%%%%%%%%%%%%%%%%%%%%%%%%%%%%%%%
\section{Introduction}
%%%%%%%%%%%%%%%%%%%%%%%%%%%%%%%%%%%%%%%%%%%%%%%%%%%%%%%%%%%%%%%%%%%%%

The top quark's large mass of order the electroweak scale suggests it may have a significant coupling to new physics in the sector which breaks electroweak symmetry.   
This makes top quark physics a natural place to test the consistency of the Standard Model (SM) and to search for effects of new physics.  
Experiments at the Tevatron and the LHC have targeted several properties of top quarks, such as the single and pair production  cross sections, pair production asymmetries, polarization, spin correlations.  
So far only one of these measurements --  the forward-backward asymmetry of \ttbar\ pair production at the Tevatron $\afbtt$ -- has shown a significant deviation from the SM.
Both  CDF   \cite{cdf_attfbcombined} and  \dzero\  \cite{Abazov:2011rq} collaborations report inclusive $\afbtt$ with central values of order $20\%$ and more than $2\sigma$ away from the prediction based on next-to-leading order (NLO) QCD \cite{KuhnandRodrigo,Bowen:2005ap}.    
Moreover, CDF (though not  \dzero\ \cite{Abazov:2011rq}) observes a strong dependence of $\afbtt$ on the invariant mass of the top pair \cite{Aaltonen:2011kc}.   
The  forward-backward asymmetry of the charged lepton in semileptonic  \cite{Abazov:2011rq} and dileptonic  \cite{cdf_attfbll} \ttbar\ events is also larger than the SM prediction.   
It is intriguing that despite noticeable differences between the individual measurements they show a large positive asymmetry that cannot be accounted for by known SM processes.\footnote{There is still room for improving the precision of SM predictions, especially concerning the effects of experimental cuts on the measured asymmetry.  For recent discussions see \cite{Kuhn:2011ri}.}
Generating sizable $\afbtt$ requires the presence of a new particle with chiral couplings not only to the top but also to the light (up and/or down) quarks. 
This implies that models which generate large $\afbtt$ also predict that observables related to top quark polarization will be affected  \cite{Godbole:2010kr,Degrande:2010kt,Choudhury:2010cd,Krohn:2011tw}. 

In this paper we point out that it is interesting to investigate the forward-backward asymmetry of the charged lepton in $q \bar q \to t \bar t$ events {\em near the production threshold}. 
This observable has a simple and intuitive theoretical interpretation, namely it  measures the relative contribution of $q_R \bar q_R$ and $q_L \bar q_L$ to top pair production at threshold. 
The reason is that  at threshold the top pair has no orbital angular momentum, therefore the top spins are determined by the chiralities of the quarks which produced them.
As usual, the top spins can be statistically measured by looking at the angular distribution of the charged lepton, which is preferentially emitted along the top quark spin and oppositely to the anti-top quark spin.\footnote{The analyzing power of top spins has long been understood, see for example \cite{Kuhn:1981md}. 
%contains a nice discussion of lepton distributions from the polarization of chiral top quarks which would occur if top quarks were light enough to be produced in $W$ decays.
} 
At tree level the SM predicts equal contributions from $q_R \bar q_R$ and $q_L \bar q_L$ to \ttbar\ production and therefore a vanishing threshold lepton asymmetry.
On the other hand, new physics models predicting large $\afbtt$ always involve different couplings to left- and right-handed light quarks (for a review and references see \cite{AguilarSaavedra:2011ug,Kamenik:2011wt}), and therefore predict positive or negative values for the threshold lepton asymmetry. 
We argue that in many cases the threshold lepton asymmetry offers a stronger discriminating power than previously considered measures of top polarization. 
From the experimental point of view the measurement is relatively straightforward, especially at the Tevatron where $q \bar q$ is the dominant top production mode and where a large fraction of the \ttbar\ pairs are produced close to threshold, $v_{top}\ll c$.
Unlike the measurement of spin correlations, the lepton asymmetry only requires looking at a single lepton independently of the rest of the event. Therefore it can be measured in semileptonic as well as in dileptonic top events.
Given the size of the top sample accumulated by the Tevatron, the threshold lepton asymmetry can be measured with a reasonably small statistical error.

%%%%%%%%%%%%%%%%%%%%%%%%%%%%%%%%%%%%%%%%%%%%%%%%%%%%%%%%%%%%%%%%%%%%%
\section{The argument}
\label{sec:spins}
%%%%%%%%%%%%%%%%%%%%%%%%%%%%%%%%%%%%%%%%%%%%%%%%%%%%%%%%%%%%%%%%%%%%%. 

Consider top quark production in a collision of ultra-relativistic quarks with definite helicity or, equivalently, with definite chirality. 
In principle, there are 4 distinct chirality configurations of the quark-antiquark pair: $q_L \bar q_L , q_L \bar q_R, q_R \bar q_L, q_R \bar q_R$. 
In a given physics model only a subset of initial states may lead to \ttbar\ production. 
For example, ignoring the light quark masses, QCD produces \ttbar\ pairs only from the $q_L \bar q_L$ and $q_R \bar q_R$ initial states.

For definiteness, we first focus on the initial state $q_R \bar q_R$. A right-chirality quark has spin in the direction of its motion, i.e. positive helicity. Right-handed anti-quarks have negative helicity. Therefore, quark and antiquark have opposite helicities, and since they are moving in opposite directions, their spins are aligned. Thus, the initial state has total spin 1 and polarization in the direction of the incoming quark (we will call this direction the positive z-direction). Since the incoming quarks do not have orbital angular momentum in the z-direction, the z-component of the total angular momentum is also 1. Using angular momentum conservation, the final state must have angular momentum 1 in the positive z-direction. 
But at threshold the top quarks have no relative velocity and no orbital angular momentum; in other words, the production process proceeds through s-wave, independently of assumptions about the interactions (i.e. QCD or new physics).
It follows that the top spins must be aligned and point in the z-direction in order to equal the spin of the initial state.

The spins of top quarks can be determined statistically by measuring the direction of the decay products. 
In particular, it is well known that the charged lepton in leptonic top decays is a ``perfect'' top spin analyzer.  
For the top, the direction of the positively charged lepton follows the distribution 
\begin{equation}
\frac{1}{\Gamma}\frac{d\Gamma}{d\cos\theta}= \frac12(1+ \cos\theta),
\label{eq:spinanalyser}
\end{equation}
where $\theta$ is the angle between the momentum of the outgoing lepton and the top spin in a reference frame where the top quark is at rest. 
For the anti-top the situation is reversed: the negative lepton has a $1-\cos \theta$ distribution with respect to the anti-top spin. 
Thus, a process in which \ttbar\ pairs are produced at rest from an initial state with right-handed initial quark chiralities predicts a distinctive angular distribution in semileptonic or dileptonic decays: the positively charged leptons are predicted to go mostly in the positive z-direction with the distribution Eq.~(\ref{eq:spinanalyser}) whereas negatively charged leptons are emitted mostly in the negative z-direction. 
% Integrating this angular distribution, one obtains a prediction for the asymmetry for the direction of the positively charged lepton 
To quantify this effect, one can define the lepton asymmetry  
\begin{equation}
\afbl = { N_l(q_l \cos \theta_l  > 0) - N_l(q_l \cos \theta_l < 0) \over N_l(q_l \cos \theta_l > 0) +  N_l(q_l \cos \theta_l < 0) }. 
\label{eq.alfb} 
\end{equation}
Here $\theta_l$ is the angle between the lepton and the incoming quark directions. 
For the  $q_R \bar q_R \to t \bar t$ process in the \ttbar\ rest frame (where both the top and anti-top are at rest at threshold)  we  obtain  the threshold lepton asymmetry $\afbl (\sqrt{s} = 2 m_{t}) =  + 50 \%$.  

If the initial state consists of left chirality quarks, $q_L \bar q_L$, all spins are reversed. 
In this case the \ttbar\ pair at threshold has spins in the negative z-direction and  $\afbl (\sqrt{s} = 2 m_{t}) = -50\%$. 
More generally, by measuring the angular distribution of the charged leptons in semileptonic or dileptonic top decays one can determine what fraction of \ttbar\ events at threshold originated from left or right-handed initial quarks.  
Of course, QCD is parity symmetric and predicts equal admixture of left- and right-chiral initial quarks. 
Thus, the QCD prediction is $\afbl (\sqrt{s} = 2 m_{t}) = 0$ at the tree-level (in fact, this holds for arbitrary $\sqrt s$).
If, however, there are new physics contributions to \ttbar\ production for which the couplings to left- and right-chiral fields differ one expects a non-vanishing lepton asymmetry. % Specifically, if the new physics couples primarily to right-handed light quarks we predict a positive lepton asymmetry for \ttbar\ threshold events whereas for models with couplings to left-handed light quarks we predict a negative asymmetry. 

Several comments are in order. 
\begin{itemize}
\item
The threshold lepton asymmetry is independent of the \ttbar\ forward-backward asymmetry $\afbtt$. 
%Actually,  $A_{FB}^{tt}$ typically vanishes at the threshold,
In fact,  it is easy to construct models in which the inclusive $\afbtt$ and $\afbl$ have opposite signs.
The threshold lepton asymmetry is also distinct from  spin correlations. 
The latter are sensitive to the {\em relative} directions of the spins of top and anti-top and at threshold it takes the same value (+1) regardless whether the top pair is produced by $q_R \bar q_R$ or $q_L \bar q_L$.  
At a practical level note that the spin correlation requires the decay products of {\em both} top and anti-top as an input, whereas in the case of the lepton asymmetry it is sufficient to look only at the lepton from either the top or anti-top. In particular, the lepton asymmetry can be measured in semileptonic events.Finally, there are important differences between the {\em threshold} and {\em inclusive} lepton asymmetries.  
As we discussed, the former depends only on the chirality of the initial quarks, while the latter is also sensitive to the chiralities and  the forward-backward asymmetry of the final state tops.   
Again, there exist models where the threshold and inclusive asymmetries have opposite signs.
\item Apart from the lepton asymmetry it is interesting to study the {\em dilepton asymmetry} which can be defined in events in which both top quarks decay leptonically
\begin{equation}
\afbll = { N(\eta_{\ell_+} >\eta_{\ell_-} ) - N(\eta_{\ell_+} < \eta_{\ell_-} )  \over  N(\eta_{\ell_+} >\eta_{\ell_-} ) + N(\eta_{\ell_+} < \eta_{\ell_-} )  },  
\label{eq.adlfb} 
\end{equation}
where $\eta$ is the pseudorapidity.      
At the \ttbar\ threshold, the dilepton asymmetry, much as the single lepton one, directly measures the initial quark helicities.
In particular, for a purely $q_R \bar q_R$ initial state it takes the value  $+2/3$, while for  $q_L \bar q_L$ it is $-2/3$, independently of other details of the production process. 
Thus the threshold dilepton  asymmetry is an even more sensitive probe of light quark polarization than the single lepton one, at least in theory.  
Moreover, the dilepton asymmetry has the advantage of being invariant under longitudinal boosts. 
On the other hand, the disadvantages of the dilepton observable are a more challenging \ttbar\ invariant mass reconstruction and smaller statistics.  
\item
While the argument given above applies at the strict \ttbar\ threshold it is clear that corrections to our analysis scale like a model-dependent power of $v_{\rm top}$.
 As long as the top quarks are non-relativistic we expect the correlation of the top spins with the beam direction to persist. We will present numerical examples which quantify  this in the next section. 
\item The analysis presented here is only valid at leading order and higher order corrections, for instance due to emission of extra gluons, would change our analytic results. However, we do not expect these effects to change the qualitative argument discussed above. Next to leading order corrections to the spin analyzing power of the lepton daughter or the spin-spin correlation coefficient were recently discussed in~\cite{Bernreuther:2010ny} and found to be important but still subdominant. 
\item The spin effect discussed here does not depend on the chirality of the top quarks produced. In fact, since the top quarks are at rest their chirality is not a good quantum number, only the top spins matter and these are completely determined by the initial quarks.
\item
The size of the effect predicted here is diluted 
by production of \ttbar\ pairs from initial state gluons, which clearly does not lead to any asymmetry. 
At the Tevatron this is not expected to be a big effect as within the SM 80\% of of the top production is from initial quark anti-quark annihilation. 
At the LHC, due to the domination of the gluon contribution and the symmetric initial state, more complicated extraction techniques are needed \cite{LHCAFB}. 
\item
Events initiated by $q_L \bar q_R$ or $q_R \bar q_L$ pairs would lead to tops with zero total spin at threshold, and thus to a vanishing threshold lepton asymmetry, an important information by itself.
We note that in this situation the spin correlation observable has opposite sign compared to the $q_L \bar q_L$ and $q_R \bar q_R$ initiated events. 
\item 
Near threshold, one expects the new physics contributions to be subdominant so that the largest effect will come from interference with standard model processes. 
In the SM the dominant production is induced from up quarks. 
Thus, the threshold lepton asymmetry is most sensitive to new physics with two structures~\cite{Delaunay:2011gv}:  
$\bar u \gamma^\mu T^a P_{L,R} u$, where $u$ is the up quark field, $\gamma^\mu$ is a Dirac matrix, $T^a$ is an SU(3) generator and $P_{L/R}$ is a projection operator. 
\end{itemize}

%%%%%%%%%%%%%%%%%%%%%%%%%%%%%%%%%%%%%%%%%%%%%%%%
\section{Example models}
%%%%%%%%%%%%%%%%%%%%%%%%%%%%%%%%%%%%%%%%%%%%%%%%

\subsection{Toy model, chiral QCD}
%%%%%%%%%%%%%%%%%%%%%%%%%%%%%%%%%%%%%%%%%%%%%%%%
To illustrate our point, let us start with a toy model where the spin effects that were discussed so far can be clearly isolated.  
The model is a chiral version of QCD where a massless {\em chiral gluon} has arbitrary couplings to left- and right-handed quarks, 
\begin{equation} 
{\cal L} \subset A_\mu^a \big(g_{q_L} \bar q \bar \sigma^\mu q  + g_{q_R}  q^c \sigma^\mu \bar q^c     + g_{t_L} \bar t \bar \sigma^\mu t  + g_{t_R}  t^c \sigma^\mu \bar t^c\big). 
\end{equation}
Consider the top production process from a pair of light quarks colliding at a fixed CM energy $\sqrt{s}$. 
The amplitude can be written as 
\beq
{\cal M}(q_i \bar q_j \to t_k \bar t_l)  = \left (\delta_{i k} \delta_{jl} - {1 \over 3}  \delta_{ij}  \delta_{kl} \right )  F(s_q,s_{\bar q}|s_t,s_{\bar t}),
\eeq 
where $i\dots l$ are color indices and  $s_x$ are the quark spins, $(+/-)$ for spin in the positive/negative z-direction. 
At {\em threshold}, $\sqrt s = 2 m_t $, we find   
\begin{eqnarray}
F(+,+|+,+)  & = & -  {g_{q_R} ( g_{t_L}  + g_{t_R} ) \over 2}, 
\nonumber \\
F(-,-|-,-)  & = & -  { g_{q_L} ( g_{t_L}  + g_{t_R} ) \over 2}, 
\end{eqnarray}
while for other spin configurations the amplitude vanishes at threshold. 
In agreement with our previous discussion,  at threshold the spin state of the top pair equals that of the incoming quark, thus, the top spins probe the coupling of the chiral gluon mediator to the light quarks. 
For example, if the chiral gluon couples only to right-handed light quarks, $g_{q_L} =0$, $g_{q_R} \neq 0$,  then both the top and the anti-top quark end up with the spin $+1/2$.  In that situation the threshold lepton asymmetry will take the maximal value of $+50\%$.
Conversely,  for $g_{q_L} \neq 0$ and $g_{q_R} = 0$ the threshold lepton asymmetry will be $-50\%$.  

\begin{figure} 
\hspace{-1cm}
\includegraphics[width=0.45\textwidth]{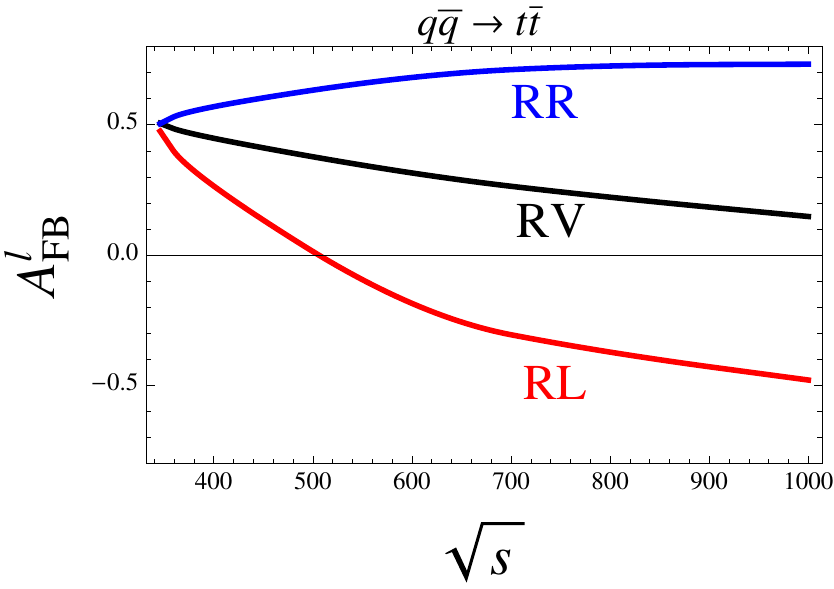}
\quad
\includegraphics[width=0.45\textwidth]{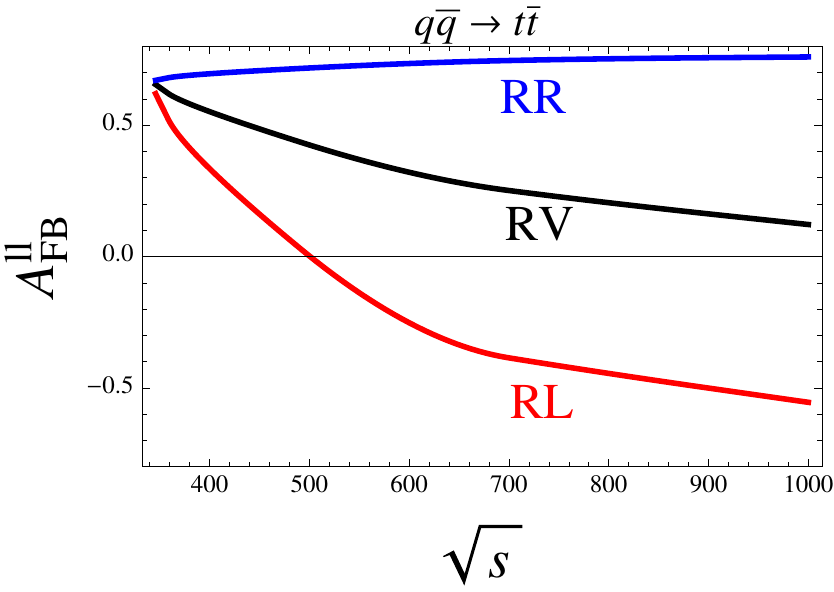}
\caption{Lepton (left) and dilepton (right) forward-backward asymmetry in the $q \bar q \to t \bar t$ process as a function of the center-of-mass production energy in the chiral-QCD toy model. 
We assumed purely right-handed couplings of the chiral gluon to the light quarks,  and  left-handed (RL), right-handed (RR) and vector (RV) couplings to the top quark.
In all three cases the  threshold lepton asymmetry is $+50\%$ and the threshold dilepton asymmetry is $+66\%$.}
\label{fig.lrv}
\end{figure}

In Fig.~\ref{fig.lrv} we plot  the lepton asymmetry  as the function of the center-of-mass \ttbar\  production energy  for purely right-handed coupling of the chiral gluon to the light quarks, 
and for 3 different couplings to the top quarks.     
We see that at thershold the lepton asymmetry is indeed always $\sim 50\%$, independently of the coupling to the top.\footnote{For axial couplings to the top the leading order amplitude vanishes at threshold. In that case the production cross-section at threshold is dominated by higher order terms in $v_{\rm top}$ expansion of the amplitude, and $\afbl$ ends up being less than $50\%$. Higher order QCD corrections, not discussed in this work, could also alter this result.}  
At higher $\sqrt s$, the \ttbar\ final state has orbital angular momentum and there is no longer a one-to-one correspondence between the spins of the light quarks and of the tops, leading to $\afbl$ different from $50\%$. 
In fact, at very high $\sqrt s$ when the tops have large velocities, the lepton asymmetry approaches the value of the top forward-backward asymmetry  which is positive (negative) for right-handed (left-handed) coupling of the chiral gluon to the top, and which is zero if that coupling is vector-like.  
The dilepton asymmetry has a very similar qualitative behavior, the main difference being the threshold value of  $\sim 66\%$. 

We stress again that the lepton forward-backward asymmetry at threshold is independent of the top forward-backward asymmetry, in particular for $g_{t_L} = g_{t_R}$ the latter is zero, while the former can be anywhere between $-50\%$ and $50\%$ depending on the relative magnitude of $g_{q_L}$ and $g_{q_R}$.  
Furthermore, we underline the difference between the threshold and inclusive lepton asymmetries.  
In our example with  $g_{q_L} = g_{t_R} = 0$ (marked as RL in the plot) the inclusive lepton asymmetry would be very small after convoluting with Tevatron PDFs,  while the threshold lepton asymmetry is $\sim 50\%$.

\subsection{Some Realistic Models}
%%%%%%%%%%%%%%%%%%%%%%%%%%%%%%%%%%%%%%%%%%%%%%
So far we have discussed a simple toy model where the contributions to the threshold lepton asymmetry are easily understood and calculated.  
Going to realistic new physics models, several new effects may complicate the picture.   
First of all, there will be the QCD contribution to the \ttbar\ production, which  is always symmetric. 
Thus, the threshold lepton asymmetry is not expected to be maximal, unlike in the toy model, unless the new physics completely dominates the \ttbar\ production at threshold  (which is unlikely given experimental constraints). 
Typically, new physics will interfere with QCD changing the relative contributions of $q_L\bar q_L$ and $q_R \bar q_R$ to the top production rate, leading to a moderate positive or negative threshold lepton asymmetry.  
Moreover, there is always the gluon initial state contribution to the \ttbar\ production which cannot favor any polarization direction. 
Of course, any effect present at the parton level is expected to be further smeared out  by detector and reconstruction effects.   
In the same vein, the single lepton asymmetry measured in the LAB frame will be smeared out after PDFs are taken into account due to the random boost of the \ttbar\ system, while the measurement in the \ttbar\ rest frame implies additional reconstruction uncertainties.   
Finally, focusing on the threshold asymmetry inevitably increases statistical uncertainties, as only a fraction of \ttbar\ pairs are produced with small momentum.  

Nevertheless, the effect  discussed above should be observable in physically interesting models, 
in particular, in many models addressing the anomalous \ttbar\ forward-backward asymmetry reported by CDF and  \dzero. 
 To demonstrate it, we study lepton asymmetry in  several models which have been proposed in the literature.
Quite generally, these models predict new particles with sizable chiral couplings to quarks which contribute to top production in s-channel \cite{Djouadi:2009nb,Tavares:2011zg}, t/u-channel \cite{Jung:2009jz}, or both \cite{Ligeti:2011vt}. For our numerical study we pick $5$ benchmark points and show the predicted lepton asymmetries near threshold in Table~\ref{t.afbaxi}. 
 Three of these benchmarks belong to a model of a light  ($m_{G'} \lesssim 2 m_t$) s-channel color octet, an ``axigluon",  having  flavor universal coupling to quarks. 
 This model is capable of producing large enough contributions to  $\afbtt$  without violating the constraints from other Tevatron and LHC observables as long as the axigluon mass is not too far above the \ttbar\ threshold and has a large width \cite{Tavares:2011zg}.  
We choose $m_{G'} = 200$ GeV, $\Gamma_{G'} = 50$ GeV and  flavor and isospin universal  axigluon couplings to the quarks as $g_R = 0.8 g_s$, $g_{L} = 0 $  (AxR),  $g_R = 0$, $g_{L} = 0.8 g_s $  (AxL), and  $g_R = -g_{L} = 0.4 g_s$  (AxA). 
Each variant of the model predicts  $\Delta \afbtt \approx 11\%$\footnote{Throughout we refer to the longitudinal boost invariant \ttbar\ asymmetry defined as $\afbtt = { N(\eta_{t} >\eta_{\bar t} ) - N(\eta_{t} < \eta_{\bar t} )  \over  N(\eta_{t} >\eta_{\bar t} ) + N(\eta_{t} < \eta_{\bar t} )  }$.}. 
 Looking at the table it is clear that the sign of the lepton asymmetry at threshold reflects the polarization of the light quarks that couple to the axigluon. 
This effect carries over to the inclusive lepton asymmetry,  as noticed in \cite{Krohn:2011tw}, however in this case the discrimination between different couplings is weaker due to a stronger correlation of 
$\afbl$ and $\afbtt$ away from the production threshold. 
Quantitatively similar results would be obtained in a model of a heavy  ($m_{G'} \gg 2 m_t$) axigluon with flavor non-universal couplings \cite{Djouadi:2009nb}. 
We also studied a model featuring an electroweak doublet scalar $S$ coupled  to the quarks via the flavor-violating Yukawa couplings $y_R S Q_3 u^c  + y_L S Q_1  t^c $  \cite{AguilarSaavedra:2011ug,Blum:2011fa}. 
In this model the scalar contributes to the $u \bar u \to t \bar t $ process in the t-channel, and for a light enough mass it produces a positive contribution to $\afbtt$. 
For our benchmark points we choose $m_S = 170$ GeV and, $y_R = 1.5$, $y_L =0$  (SdR) or $y_R = 0$, $y_L = 1.5$  (SdL), which leads to $\Delta \afbtt \approx 6\%$.  
Here, somewhat counterintuitively, the lepton asymmetry at threshold is negative (positive) when  the scalar couples to right-handed (left-handed) up quarks.   
What happens is that the scalar t-channel exchange interferes {\em destructively} with QCD, thus,  for moderate $y_R$ $(y_L)$ couplings, the contribution of the right-handed (left-handed) quarks to the top production rate gets {\em suppressed}.

\begin{table}
\begin{tabular}{|c|c|c|c|c|c|} 
\hline 
Benchmark &  $\sqrt s < 375 \gev $  & $\sqrt s < 450 \gev $ &  Inclusive
\\ \hline 
AxR &   4\bs 13/18/21\%   & 8\bs 14/17/21\%  &  11\bs 14/17/21\%   \\ \hline 
AxL &  5\bs -10/-13/-18\% &  9\bs -8/-10/-13\%  &  11\bs -6/-7/-9\% \\ \hline  
AxA &   6\bs 2/2/2\% & 10\bs 3/3/5\% & 12\bs 5/6/7\%     \\ \hline 
SdR &  0\bs - 7/-10/-13\% &    2\bs -6/-7/-9\%  &  5\bs -3/-4/-4\%   \\ \hline 
SdL &  2\bs  8/10/14 \% &    3\bs 8/10/12\%  &  6\bs 9/11/13\%   \\ \hline  \hline 
Fraction & 17\%  & 60\% &100\% 
\\ \hline 
\end{tabular}
\caption{The predicted Tevatron lepton forward-backward asymmetry calculated using MadGraph 5 \cite{Alwall:2011uj}  for several benchmarks defined in the text. 
 The asymmetry is integrated for 3 different $m_{tt}$ intervals from threshold up. 
In each entry, we give the 4 observables $\afbtt$ \bs\, $\afbl$(LAB frame)/ $\afbl$(\ttbar\ rest frame) / $\afbll$.
All numbers are fully inclusive (no cuts), parton level (no showering and detector effects), and tree-level (in particular, purely SM NLO contributions are not included).   
The last line in the table gives the approximate fraction of the total \ttbar\ rate falling into the corresponding $m_{tt}$ interval. % (these fractions very weakly depend on a benchmark point).
 }
\label{t.afbaxi}
\end{table}
  
%%%%%%%%%%%%%%%%%%%%%%%%%%%%%
\section{Existing related data} %and relation with other observables}
%%%%%%%%%%%%%%%%%%%%%%%%%%%%%

The threshold lepton asymmetry has not been tackled experimentally so far,   
but there are two measurements  of the inclusive lepton asymmetry in \ttbar\ events.  
In Ref. \cite{Abazov:2011rq}  \dzero\ reports the measurement of the single lepton  asymmetry  in semileptonic top  events. %  using 5.4~fb$^{-1}$ of data.
At the production (parton) level the result is   
\beq
\afbl  = (15.2 \pm 4.0) \%,
\eeq
compared to $2\%$ predicted by the SM at the NLO.
Interestingly, this measurements' departure from the SM prediction is statistically more significant than the deviation in the \ttbar~asymmetry measured in the same sample: 
$\afbtt= 19.6 \pm 6.5 \%$.  

The other result is thanks to the CDF collaboration \cite{cdf_attfbll} who measured the lepton asymmetry in dileptonic  top events.
At the reconstruction level after subtracting non-\ttbar\ background CDF obtains 
\beq
\afbl  = (21 \pm 7) \%, 
\eeq
which is similar to the measured \ttbar~asymmetry at the reconstruction level and strikingly smaller than the parton level $\afbtt = (42 \pm 15)\%$ in that channel \cite{cdf_attfbll}.\footnote{Note, however, that due to the rather hard leptonic $p_T$ cuts and missing $E_T$ applied in this analysis it should not be considered a fully inclusive asymmetry measurement.} 
Finally, recall that  combining the semileptonic and dileptonic channels at CDF one finds the inclusive parton level  $\afbtt = (20 \pm 7)\%$  \cite{cdf_attfbcombined}, very close to the  \dzero\  value. 
Thus, both collaborations find that the inclusive lepton asymmetry has the same sign as the inclusive \ttbar~asymmetry which provides a handle  for discriminating between different new physics models. % \cite{Krohn:2011tw}. 
In fact, the published  \dzero\  and CDF measurements already strongly disfavor models predicting negative inclusive lepton asymmetries, such as our AxL and SdR benchmarks in Table~\ref{t.afbaxi}.
Moreover, we observe that for both CDF and  \dzero\  the measured lepton asymmetry is smaller than the parton level \ttbar\ asymmetry. 
This fits best with the models predicting comparable contributions of $q_R \bar q_R$ and $q_L \bar q_L$ to the top production, such as our AxA benchmark, although at this point  a moderate domination of $q_R \bar q_R$ cannot be excluded. 
 Measuring the threshold lepton asymmetry would provide a valuable piece of information, allowing a better discrimination between competing models.

%%%%%%%%%%%%%%%%%%%%%%%%%%%%%%%%%%%%%%%%%%%%%%%%
\section{Conclusions and Outlook}
\label{sec:conclusion}
We have argued that lepton forward-backward asymmetry in $t\bar t$ events  at threshold carries direct information of the production mechanism. 
In particular, when top pairs are produced from purely right-handed (left-handed) quarks a positive (negative) threshold lepton asymmetry is predicted. 
To our knowledge, this simple argument has not been explicitly made in the earlier literature, and has not been explored in experimental studies.

Threshold  lepton asymmetry is independent of the top couplings and may be present even when the \ttbar\ forward-backward asymmetry vanishes. 
It may be studied for a single lepton in semileptonic or dileptonic events in which case it does not require looking at the decay products of the other top.   
One may also study  the threshold dilepton asymmetry in dileptonic events which carries the same information.   
The threshold asymmetry can be very different from the inclusive asymmetry. 
Charged leptons from tops at threshold probe the chirality of the initial quarks, 
whereas for highly boosted tops the lepton direction is correlated with the direction and chirality of the top.
The inclusive lepton asymmetry is thus a convolution of several distinct effects. 
 %as the latter correlated with the  \ttbar\ forward-backward asymmetry and the chirality of the produced top quarks.  
Furthermore, the threshold lepton asymmetry is complimentary to the spin-spin correlation measurement~\cite{Choudhury:2010cd,Degrande:2010kt}; the latter does not distinguish $q_R \bar q_R$ and $q_L \bar q_L$ initiated events but is more sensitive to an admixture of $q_R \bar q_L$ and $q_L \bar q_R$ initiated events.  
 
In this paper we studied  lepton asymmetry at the Tevatron only. 
At the LHC that asymmetry of course vanishes due to the symmetric initial state. 
However, by focusing on events in which the center of mass of the \ttbar\ pair is highly boosted one can, in principle, gain  access to the asymmetry \cite{lhcth}. 
Highly boosted events (with the \ttbar\ pair still at rest relative to each other) are much more likely to have originated from quarks than from gluons and the direction of the boost provides a statistical tag for the direction of the initial quark. The asymmetry to be measured at the LHC would then be an asymmetry of the lepton with respect to the direction of the overall longitudinal boost of the event.  
Whether the threshold lepton asymmetry can be realistically observed at the LHC is a non-trivial question which deserves further study. 
 
Note that our simulations were restricted to the parton level. 
The effects of showering and detector resolution should be taken into account, although they are not expected to be as important as in the case of \ttbar\ asymmetry.    
A separate question is the impact of NLO QCD and bound state corrections on the predictions of the threshold lepton asymmetry. 
In this paper we purposefully avoided these issues. 
However one expects lepton angular distributions to be robust against soft QCD effects. 
This has been demonstrated in the related cases of polarized $e^+ e^- \rightarrow$ \ttbar\ production \cite{Harlander:1994ac} and  $\gamma \gamma \to t \bar t$ production \cite{khoze}. 
%as we are planning to address them in future publications. However

%%%%%%%%%%%%%%%%%%%%%%%%%%%%%%%%%%%%%%%%%%%%%%%%

%%%%%%%%%%%%%%%%%%%%%%%%%%%%%%%%%%%%%%%%%%%%%%%%
\acknowledgments
%%%%%%%%%%%%%%%%%%%%%%%%%%%%%%%%%%%%%%%%%%%%%%%%

 We acknowledge useful conversations with Andy Cohen, Christophe Grojean, Tom Schwartz, Pekka Sinervo, Christian Spethmann, and Jessie Thaler. We also acknowledge the CERN ``TH-LPCC Summer Institute on LHC Physics'' during which this project was initiated.
MS was supported in part by DOE grant DE-FG02-01ER-40676. 
 
%%%%%%%%%%%%%%%%%%%%%%%%%%%%%%%%%%%%%%%%%%%%%%%%%%%%%%%%%%%%%%%%%%


\begin{thebibliography}{99}
%%%%%%%%%%%%%%%%%%%%%%%%%%%%%%%%%%%%%%%%%%%%%%%%%%%%%%%%%%%%%%%%%%

\bibitem{cdf_attfbcombined}
T. Aaltonen et al. [CDF Collaboration], CDF Conference Note 10584.

%\cite{Abazov:2011rq}
\bibitem{Abazov:2011rq}
  V.~M.~Abazov {\it et al.} [ D0 Collaboration ],
  %``Forward-backward asymmetry in top quark-antiquark production,''
    [arXiv:1107.4995 [hep-ex]].
 
 
 %\cite{KuhnandRodrigo}
\bibitem{KuhnandRodrigo}
  J.~H.~Kuhn and G.~Rodrigo,
  %``Charge asymmetry in hadroproduction of heavy quarks,''
  Phys.\ Rev.\ Lett.\  {\bf 81}, 49 (1998),
  [arXiv:hep-ph/9802268]. 
  %%CITATION = PRLTA,81,49;%%
  J.~H.~Kuhn and G.~Rodrigo,
  %``Charge asymmetry of heavy quarks at hadron colliders,''
  Phys.\ Rev.\  D {\bf 59}, 054017 (1999),
  [arXiv:hep-ph/9807420].
  %%CITATION = PHRVA,D59,054017;%%
  %\cite{Antunano:2007da}
% \bibitem{Antunano:2007da}
%
%\cite{Bowen:2005ap}
\bibitem{Bowen:2005ap}
  M.~T.~Bowen, S.~D.~Ellis, D.~Rainwater,
  %``Standard model top quark asymmetry at the Fermilab Tevatron,''
  Phys.\ Rev.\  {\bf D73}, 014008 (2006)m
  [hep-ph/0509267].
  O.~Antunano, J.~H.~Kuhn and G.~Rodrigo,
  %``Top quarks, axigluons and charge asymmetries at hadron colliders,''
  Phys.\ Rev.\  D {\bf 77}, 014003 (2008),
  [arXiv:0709.1652 [hep-ph]].
  %%CITATION = PHRVA,D77,014003;%%
%\cite{Almeida:2008ug}
%\bibitem{Almeida:2008ug}
  L.~G.~Almeida, G.~F.~Sterman, W.~Vogelsang,
  %``Threshold Resummation for the Top Quark Charge Asymmetry,''
  Phys.\ Rev.\  {\bf D78}, 014008 (2008),
  [arXiv:0805.1885 [hep-ph]].
    
%\cite{Aaltonen:2011kc}
\bibitem{Aaltonen:2011kc}
  T.~Aaltonen {\it et al.}  [CDF Collaboration],
  %``Evidence for a Mass Dependent Forward-Backward Asymmetry in Top Quark Pair
  %Production,''
  arXiv:1101.0034 [hep-ex].
  %%CITATION = ARXIV:1101.0034;%%

\bibitem{cdf_attfbll}
T. Aaltonen et al. [CDF Collaboration], CDF Conference Note 10436.


%\cite{Kuhn:2011ri}
\bibitem{Kuhn:2011ri}
%\cite{Kidonakis:2011ca}
%\bibitem{Kidonakis:2011ca}
  N.~Kidonakis, B.~D.~Pecjak,
  %``Top-quark production and QCD,''
  %Submitted to: Eur.Phys.J.C.
  [arXiv:1108.6063 [hep-ph]].
  J.~H.~Kuhn, G.~Rodrigo,
  %``Charge asymmetries of top quarks at hadron colliders revisited,''
    [arXiv:1109.6830 [hep-ph]].


%\cite{Godbole:2010kr}
\bibitem{Godbole:2010kr}
  R.~M.~Godbole, K.~Rao, S.~D.~Rindani, R.~K.~Singh,
  %``On measurement of top polarization as a probe of $t \bar t$ production mechanisms at the LHC,''
  JHEP {\bf 1011}, 144 (2010), 
  [arXiv:1010.1458 [hep-ph]]. 
%\bibitem{Jung:2010yn}
  D.~-W.~Jung, P.~Ko, J.~S.~Lee,
  %``Longitudinal top polarization as a probe of a possible origin of forward-backward asymmetry of the top quark at the Tevatron,''
  Phys.\ Lett.\  {\bf B701}, 248-254 (2011),
  [arXiv:1011.5976 [hep-ph]].
%\cite{Cao:2010nw}
%\bibitem{Cao:2010nw}
  J.~Cao, L.~Wu, J.~M.~Yang,
  %``New physics effects on top quark spin correlation and polarization at the LHC: a comparative study in different models,''
  Phys.\ Rev.\  {\bf D83}, 034024 (2011),
  [arXiv:1011.5564 [hep-ph]].
  %\cite{Barger:2011ya}
%\bibitem{Barger:2011ya}
  V.~Barger, W.~-Y.~Keung, C.~-T.~Yu,
  %``Parity Nonconservation in Strong Interactions,'' 
  [arXiv:1108.2275 [hep-ph]].

%\cite{Degrande:2010kt}
 \bibitem{Degrande:2010kt}
  C.~Degrande, J.~-M.~Gerard, C.~Grojean, F.~Maltoni, G.~Servant,
  %``Non-resonant New Physics in Top Pair Production at Hadron Colliders,''
  JHEP {\bf 1103}, 125 (2011), 
  [arXiv:1010.6304 [hep-ph]].

\bibitem{Choudhury:2010cd}
  D.~Choudhury, R.~M.~Godbole, S.~D.~Rindani, P.~Saha,
  %``Top polarization, forward-backward asymmetry and new physics,''
    [arXiv:1012.4750 [hep-ph]].

\bibitem{Krohn:2011tw}
  D.~Krohn, T.~Liu, J.~Shelton, L.~-T.~Wang,
  %``A Polarized View of the Top Asymmetry,''
    [arXiv:1105.3743 [hep-ph]].

%\cite{Kuhn:1981md}
\bibitem{Kuhn:1981md}
  J.~H.~Kuhn, K.~H.~Streng,
  %``Measurement Of Weak Couplings Through Toponium Decays,''
  Nucl.\ Phys.\  {\bf B198}, 71 (1982).
%\bibitem{Kuhn:1983ix}
  J.~H.~Kuhn,
  %``How to Measure the Polarization of Top Quarks,''
  Nucl.\ Phys.\  {\bf B237}, 77 (1984).


%\cite{AguilarSaavedra:2011ug}
\bibitem{AguilarSaavedra:2011ug}
  J.~A.~Aguilar-Saavedra, M.~Perez-Victoria,
  %``Simple models for the top asymmetry: Constraints and predictions,''
    [arXiv:1107.0841 [hep-ph]].
%\cite{Kamenik:2011wt}
\bibitem{Kamenik:2011wt}
J.~F.~Kamenik, J.~Shu, J.~Zupan,
  %``Review of new physics effects in t-tbar production,''
  [arXiv:1107.5257 [hep-ph]].
%\cite{Westhoff:2011tq}
%\bibitem{Westhoff:2011tq}
  S.~Westhoff,
  %``Top-Quark Asymmetry -- A New Physics Overview,''
  [arXiv:1108.3341 [hep-ph]].
  
%\cite{Bernreuther:2010ny}
\bibitem{Bernreuther:2010ny}
  W.~Bernreuther, Z.~-G.~Si,
  %``Distributions and correlations for top quark pair production and decay at the Tevatron and LHC.,''
  Nucl.\ Phys.\  {\bf B837}, 90-121 (2010),
  [arXiv:1003.3926 [hep-ph]].

\bibitem{LHCAFB}
[CMS Collaboration], CMS-PAS-TOP-10-010, CMS-PAS-TOP-11-014.
[ATLAS Collaboration], ATLAS-CONF-2011-106. 

\bibitem{Delaunay:2011gv}
  C.~Delaunay, O.~Gedalia, Y.~Hochberg, G.~Perez, Y.~Soreq,
  %``Implications of the CDF $t \bar{t}$ Forward-Backward Asymmetry for Hard Top Physics,''
  JHEP {\bf 1108}, 031 (2011)
  [arXiv:1103.2297 [hep-ph]].
 K.~Blum, C.~Delaunay, O.~Gedalia, Y.~Hochberg, S.~J.~Lee, Y.~Nir, G.~Perez, Y.~Soreq,
  %``Implications of the CDF $t\bar{t}$ Forward-Backward Asymmetry for Boosted Top Physics,''
  Phys.\ Lett.\  {\bf B702}, 364-369 (2011),
  [arXiv:1102.3133 [hep-ph]].


%%%%%%%%%%%%%%%%%%%%
% Heavy  axigluon 
%%%%%%%%%%%%%%%%%%%%
%\cite{Djouadi:2009nb}
\bibitem{Djouadi:2009nb}
  A.~Djouadi, G.~Moreau, F.~Richard, R.~K.~Singh,
  %``The Forward-backward asymmetry of top quark production at the Tevatron in warped extra dimensional models,''
  Phys.\ Rev.\  {\bf D82}, 071702 (2010),
  [arXiv:0906.0604 [hep-ph]].
  %\cite{Frampton:2009rk}
%\bibitem{Frampton:2009rk}
  P.~H.~Frampton, J.~Shu, K.~Wang,
  %``Axigluon as Possible Explanation for p anti-p ---> t anti-t Forward-Backward Asymmetry,''
  Phys.\ Lett.\  {\bf B683}, 294-297 (2010),
  [arXiv:0911.2955 [hep-ph]].
  %\cite{Bauer:2010iq}
%\bibitem{Bauer:2010iq}
  M.~Bauer, F.~Goertz, U.~Haisch, T.~Pfoh, S.~Westhoff,
  %``Top-Quark Forward-Backward Asymmetry in Randall-Sundrum Models Beyond the Leading Order,''
  JHEP {\bf 1011}, 039 (2010),
  [arXiv:1008.0742 [hep-ph]].
%\bibitem{Delaunay:2011vv}
  C.~Delaunay, O.~Gedalia, S.~J.~Lee, G.~Perez, E.~Ponton,
  %``Extraordinary Phenomenology from Warped Flavor Triviality,''
  Phys.\ Lett.\  {\bf B703}, 486-490 (2011),
  [arXiv:1101.2902 [hep-ph]].
  %\cite{Bai:2011ed}
%\bibitem{Bai:2011ed}
  Y.~Bai, J.~L.~Hewett, J.~Kaplan, T.~G.~Rizzo,
  %``LHC Predictions from a Tevatron Anomaly in the Top Quark Forward-Backward Asymmetry,''
  [arXiv:1101.5203 [hep-ph]].
%\bibitem{Djouadi:2011aj}
  A.~Djouadi, G.~Moreau, F.~Richard,
  %``Forward-backward asymmetries of the bottom and top quarks in warped extra-dimensional models: LHC predictions from the LEP and Tevatron anomalies,''
  Phys.\ Lett.\  {\bf B701}, 458-464 (2011),
  [arXiv:1105.3158 [hep-ph]].
%\cite{Barcelo:2011fw}
%\bibitem{Barcelo:2011fw}
  R.~Barcelo, A.~Carmona, M.~Masip, J.~Santiago,
  %``Gluon excitations in t tbar production at hadron colliders,''
  Phys.\ Rev.\  {\bf D84}, 014024 (2011),
  [arXiv:1105.3333 [hep-ph]].
%\cite{Haisch:2011up}
%\bibitem{Haisch:2011up}
  U.~Haisch, S.~Westhoff,
  %``Massive Color-Octet Bosons: Bounds on Effects in Top-Quark Pair Production,''
  JHEP {\bf 1108}, 088 (2011),
  [arXiv:1106.0529 [hep-ph]].
%\cite{Barcelo:2011vk}
%\bibitem{Barcelo:2011vk}
  R.~Barcelo, A.~Carmona, M.~Masip, J.~Santiago,
  %``Stealth gluons at hadron colliders,''
    [arXiv:1106.4054 [hep-ph]].


%%%%%%%%%%%%%%%%%%%%
% Light axigluon 
%%%%%%%%%%%%%%%%%%%%
%\cite{Tavares:2011zg}
\bibitem{Tavares:2011zg}
  G.~M.~Tavares, M.~Schmaltz,
  %``Explaining the t-tbar asymmetry with a light axigluon,''
    [arXiv:1107.0978 [hep-ph]].
%\cite{AguilarSaavedra:2011ci}
%\bibitem{AguilarSaavedra:2011ci}
  J.~A.~Aguilar-Saavedra, M.~Perez-Victoria,
  %``Shaping the top asymmetry,''
  [arXiv:1107.2120 [hep-ph]].
%\cite{Krnjaic:2011ub}
%\bibitem{Krnjaic:2011ub}
  G.~Z.~Krnjaic,
  %``Very Light Axigluons and the Top Asymmetry,''
  [arXiv:1109.0648 [hep-ph]].


%\cite{Alwall:2011uj}
\bibitem{Alwall:2011uj}
  J.~Alwall, M.~Herquet, F.~Maltoni, O.~Mattelaer, T.~Stelzer,
  %``MadGraph 5 : Going Beyond,''
  JHEP {\bf 1106}, 128 (2011).
  [arXiv:1106.0522 [hep-ph]].

%%%%%%%%%%%%%%%%%%%%
% t-channel models 
%%%%%%%%%%%%%%%%%%%%

%\cite{Jung:2009jz}
\bibitem{Jung:2009jz}
%\cite{Shu:2009xf}
%\bibitem{Shu:2009xf}
  J.~Shu, T.~M.~P.~Tait, K.~Wang,
  %``Explorations of the Top Quark Forward-Backward Asymmetry at the Tevatron,''
  Phys.\ Rev.\  {\bf D81}, 034012 (2010),
  [arXiv:0911.3237 [hep-ph]].
  %
  S.~Jung, H.~Murayama, A.~Pierce, J.~D.~Wells,
  %``Top quark forward-backward asymmetry from new t-channel physics,''
  Phys.\ Rev.\  {\bf D81}, 015004 (2010),
  [arXiv:0907.4112 [hep-ph]].
%\cite{Shelton:2011hq}
%\bibitem{Shelton:2011hq}
  J.~Shelton, K.~M.~Zurek,
  %``Maximal flavor violation from new right-handed gauge bosons,''
  Phys.\ Rev.\  {\bf D83}, 091701 (2011),
  [arXiv:1101.5392 [hep-ph]].
%\cite{Barger:2010mw}
%\bibitem{Barger:2010mw}
  V.~Barger, W.~-Y.~Keung, C.~-T.~Yu,
  %``Asymmetric Left-Right Model and the Top Pair Forward-Backward Asymmetry,''
  Phys.\ Rev.\  {\bf D81}, 113009 (2010),
  [arXiv:1002.1048 [hep-ph]].
  

%%%%%%%%%%%%%%%%%%%%
% Flavor symmetric  models 
%%%%%%%%%%%%%%%%%%%%
\bibitem{Ligeti:2011vt}
%\bibitem{Grinstein:2011yv}
  B.~Grinstein, A.~L.~Kagan, M.~Trott, J.~Zupan,
  %``Forward-backward asymmetry in t anti-t production from flavour symmetries,''
  [arXiv:1102.3374 [hep-ph]].
%\cite{Ligeti:2011vt}
  Z.~Ligeti, G.~M.~Tavares, M.~Schmaltz,
  %``Explaining the t tbar forward-backward asymmetry without dijet or flavor anomalies,''
  JHEP {\bf 1106}, 109 (2011).
  [arXiv:1103.2757 [hep-ph]].
  
  %\cite{Blum:2011fa}
\bibitem{Blum:2011fa}
 K.~Blum, Y.~Hochberg, Y.~Nir,
  %``Scalar-mediated $t\bar t$ forward-backward asymmetry,''
   [arXiv:1107.4350 [hep-ph]].


\bibitem{lhcth}
  Y.~k.~Wang, B.~Xiao and S.~h.~Zhu,
  %``One-side Forward-backward Asymmetry in Top Quark Pair Production at CERN
  %Large Hadron Collider,''
  Phys.\ Rev.\  D {\bf 82}, 094011 (2010)
  [arXiv:1008.2685 [hep-ph]].
  %\cite{Bhattacherjee:2011nr}
%\bibitem{Bhattacherjee:2011nr}
  B.~Bhattacherjee, S.~S.~Biswal, D.~Ghosh,
  %``Top quark forward-backward asymmetry at Tevatron and its implications at the LHC,''
    [arXiv:1102.0545 [hep-ph]].
    %\cite{Craig:2011an}
%\bibitem{Craig:2011an}
  N.~Craig, C.~Kilic, M.~J.~Strassler,
  %``LHC Charge Asymmetry as Constraint on Models for the Tevatron Top Anomaly,''
  Phys.\ Rev.\  {\bf D84}, 035012 (2011),
  [arXiv:1103.2127 [hep-ph]].
%\cite{Hewett:2011wz}
%\bibitem{Hewett:2011wz}
  J.~L.~Hewett, J.~Shelton, M.~Spannowsky, T.~M.~P.~Tait, M.~Takeuchi,
  %``$A^t_{FB}$ Meets LHC,''
    [arXiv:1103.4618 [hep-ph]].
     %\cite{Bai:2011uk}
%\bibitem{Bai:2011uk}
  Y.~Bai, Z.~Han,
  %``Improving the Top Quark Forward-Backward Asymmetry Measurement at the LHC,''
  [arXiv:1106.5071 [hep-ph]].
    %\cite{Gresham:2011fx}
%\bibitem{Gresham:2011fx}
  M.~I.~Gresham, I.~-W.~Kim, K.~M.~Zurek,
  %``Tevatron Top $A_{FB}$ Versus LHC Top Physics,''  
  [arXiv:1107.4364 [hep-ph]].
  
\bibitem{Harlander:1994ac} See for example 
  R.~Harlander, M.~Jezabek, J.~H.~Kuhn, T.~Teubner,
  %``Polarization in top quark pair production near threshold,''
  Phys.\ Lett.\  {\bf B346}, 137-142 (1995),
  [hep-ph/9411395].
\bibitem{khoze} See for example 
V.~S.~Fadin, V.~A.~Khoze and M.~I.~Kotsky,
 %``Top quark polarization as a probe of t anti-t  dynamics,''
 Z.\ Phys.\  C {\bf 64}, 45 (1994),
 [arXiv:hep-ph/9403246].



\end{thebibliography}
\end{document}